						[2001/04/25 1.1 (PWD)]
\documentclass[a4paper,twocolumn]{esapub} 

\usepackage{times}
\usepackage{natbib}
\usepackage{graphicx}

\begin{document}

\title{The ``Supercritical Pile" GRB MOdel: Afterglows and GRB, XRR, XRF 
Unification}
\author[1]{Demosthenes Kazanas}
\author[2]{Apostolos Mastichiadis}
\affil[1]{Astrophysics Science Division, NASA/GSFC, Greenbelt, MD 20771}
\author[1,3]{Markos Georganopoulos}
\affil[2]{Department of Astronomy, University of Athens, GR 15784, Athens, Greece}
\affil[3]{Department of Physics, JCA, UMBC, 1000 Hilltop Circle, Baltimore, MD 
21250}

\newcommand{\btx}{\textsc{Bib}\TeX}
\newcommand{\filename}{esapub}
\newcommand{\Ep}{$E_{\rm p}$}
\def\cm{\ifmmode {\rm cm}^{-1} \else cm$^{-1}$ \fi}
\def\s{\ifmmode {\rm s}^{-1} \else s$^{-1}$ \fi}
\def\cc{\ifmmode {\rm cm}^{-3} \else cm$^{-3}$ \fi}
\def\cs{\ifmmode {\rm cm}^{-2} \else cm$^{-2}$ \fi}
\def\g{\ifmmode \gamma \else $\gamma$\fi}
\def\G{\ifmmode \Gamma \else $\Gamma$\fi}
\def\Gs{\ifmmode \Gamma~ \else $\Gamma~$\fi}

\def\gc{\ifmmode \gamma_{\rm c} \else $\gamma_{\rm c}$ \fi}
\def\sw{Schwarzschild~}
\def\gsim{\mathrel{\raise.5ex\hbox{$>$}\mkern-14mu
             \lower0.6ex\hbox{$\sim$}}}

\def\lsim{\mathrel{\raise.3ex\hbox{$<$}\mkern-14mu
             \lower0.6ex\hbox{$\sim$}}}



\maketitle

\begin{abstract}
We present the general notions and observational consequences of 
the ``Supercritical Pile" GRB model; the fundamental feature of this model is a 
detailed process for the conversion of the energy stored in relativistic
protons in the GRB Relativistic Blast Waves (RBW) into relativistic 
electrons and then into radiation. The conversion is effected through the 
$p \, \gamma \rightarrow p \, e^+e^-$ reaction, whose kinematic threshold 
is imprinted on the GRB spectra to provide a peak of their emitted luminosity 
at energy \Ep $\sim 1$ MeV in the lab frame. We extend this model to include, 
in addition to the (quasi--)thermal relativistic post-shock protons an 
accelerated component of power law form. This component guarantees the production 
of $e^+e^- -$pairs even after the RBW has slowed down to the point that its 
(quasi--)thermal protons cannot fulfill the threshold of the above reaction. 
We suggest that this last condition marks the transition from the prompt to the 
afterglow GRB phase. We also discuss conditions under which this transition is 
accompanied by a significant drop in the flux and could thus account for several 
puzzling, recent observations. Finally, we indicate that the 
same mechanism applied to the late stages of the GRB evolution leads to a 
decrease in \Ep $\propto \Gamma^2(t)\propto t^{-3/4}$, a feature amenable to 
future observational tests. 

\end{abstract}

\section{Introduction}

Despite the great progress in the GRB field with the launch of {\sl CGRO} and 
the discovery of their afterglows by {\sl BeppoSAX} \citep{Costa97} (a fact that
allowed the determination of their redshifts and provided an 
unequivocal measure of their distance and luminosity), a detailed understanding
of the physics involved is still lacking. Several major issues concerning the 
dynamics and radiative processes of these events remain, to say the least,
opaque. While the models of the GRB afterglow emission are reasonably successful
in providing the time  evolution of their afterglow flux \citep{spn98, sph99},
there is an entire host of issues which are tacitly ignored while discussing 
and modeling the details of GRB emission. Some of these issues have been with 
us since the inception of the cosmological GRB models, while others are rather 
new, the outcome of the wealth of new observations, 
most prominently those by {\sl Swift} and {\sl HETE}. 

We provide below a partial list of the most important (to our view) such 
open GRB issues.

\begin{itemize}
\item The nature of the GRB ``inner engine". It is widely believed that this
has its origin to the collapse of a stellar core or the merging of two neutron 
stars (we will not be concerned with it in the rest of the paper).
\item The dissipation of the RBW kinetic energy. This is converted efficiently  
into relativistic protons (1/2000th of it into electrons) of Lorentz factor (LF)
$\sim \Gamma$ ($\Gamma$ is the LF of the blast wave). However, the efficient
 conversion of the latter into radiation, needed to account for the observed GRB
variability, is one of the least discussed or explored issues. 
\item The energy distribution of the prompt GRB phase, in particular the 
presence of a peak in the $\nu \, F_{\nu}$ spectra at \Ep$ \sim m_e c^2$ 
\citep{mallozzi95}, \citep{Preece00}.
\item The differentiation between prompt and afterglow phases. What determines, 
besides detector sensitivity, the separation in and the transition  
from the prompt to the afterglow stages in a GRB? 
\item Is there a unification between GRB, XRR and XRF? The discovery of transients
with timing properties similar to those of GRB but with smaller \Ep~ values (XRR, 
XRF) 
and smaller isotropic luminosities, $E_{\rm iso}$, raised the question of their 
relation to GRB and the a possible unification of all these phenomena. 
\end{itemize} 

\section{The ``Supercritical Pile" Model}

The model outlined in the present note has been conceived to resolve the
second of the issues enumerated above, i.e. that of the dissipation of the 
energy stored in the form of relativistic protons within the RBW of a GRB. 
The name of the model is of relevance: Its basic physics is a radiative-type 
instability which can convert the free energy of the relativistic proton 
plasma into relativistic $e^+ e^-$ pairs on time scales $R/c$ ($R$ is typical 
size of the plasma), under certain criticality conditions\citep{kim92},\citep{mak95}. 
The nickname of the model derives from the fact that 
the criticality conditions are identical to those of a supercritical nuclear pile, 
where the free energy of nuclear binding can be explosively released once these
conditions are fulfilled. 

\subsection{ The Thresholds}

The instability of a relativistic proton plasma, in distinction with that of a 
nuclear pile, involves a two step process: (a) The production of $e^+ 
e^- -$pairs by the $p \, \gamma \rightarrow p \, e^+e^-$ reaction. (b) The production of 
additional photons, through the synchrotron process, which can then serve as targets 
for the above reaction to produce more pairs etc., offering the possibility of
a runaway. Item (a) entails a {\sl kinematic} threshold, while the true criticality
is related to item (b) which in turn entails a {\sl dynamic} threshold. 

Consider a spherical volume of size $R$, containing a relativistic proton 
plasma of {\sl integral} spectrum $n_p(\g) = n_{\rm o} \g^{-\beta}$ (\g~being 
the proton Lorentz factor), along with an (infinitesimal)
number of photons of energy $\epsilon$ (in units of $m_ec^2$); 
these photons can produce pairs via the $p\g \rightarrow p \, e^-e^+$ 
reaction, provided that the proton population
extends to Lorentz factors $\g >\g_{\rm c}$ 
such that $\g_{\rm c} \, \epsilon \simeq2$. In the presence of a magnetic 
field $B$, the pairs (of Lorentz factor also equal to $\g_{\rm c}$)
produce synchrotron photons of energy $\epsilon_s = b \g_{\rm c}^2$
where $b = B/B_{\rm cr}$ is the magnetic field in units of the 
critical one $B_{\rm cr} = m_e^2 c^3/(e \hbar) \simeq 4.4 \, 10^{13}$ G.
For the reaction network to be {\sl self-contained} the energies of the seed
and synchrotron photons should be equal, yielding the {\sl kinematic}
threshold of the process i.e. 

\begin{equation}
\g_{\rm c} \epsilon_s =\g_{\rm c}^3 \, b \simeq2  ~~~{\rm or} ~~~ \g_{\rm c} \gsim
(2/b)^{1/3}.
\label{kinematic0}
\end{equation}
\
 
The reaction network will be {\sl self-sustained} if {\sl at least one} of the 
synchrotron photons pair--produces before escaping the volume of the plasma.
Since the number of photons produced by an electron of energy $\gamma$ is 
$N_{\gamma} \simeq \gamma/b\, \gamma^2 = 1/b\, \gamma$, the {\sl dynamic } 
threshold implies that the column density of the protons at energy $\g_{\rm c}$
should be greater than $1/N_{\gamma}$ or
\begin{equation}
n_{\rm o} \, \g_{\rm c}^{-\beta} \sigma_{_{\rm p \gamma}} R \gsim b \gamma
\end{equation}

where $\sigma_{_{\rm p \gamma}}$
is the cross section of the $p \, \gamma \rightarrow e^+ e^-$ reaction. 

In the case of the RBW of a GRB, on which the majority of particles are 
relativistic with mean energy  $\langle E \rangle \sim \Gamma$, we can eschew the 
presence of an accelerated non-thermal relativistic population and consider only the 
``thermal" relativistic protons present behind the forward shock of the RBW. 
These can be considered as monoenergetic of energy $\g_{_{\rm c}} = \Gamma$ (or of a 
Maxwellian of similar mean energy) where $\Gamma$ is the Lorentz factor of the RBW. 
The linear dimension of the plasma in this case should be considered to be the 
(comoving) width of the RBW $\Delta_{\rm com}$, while its density the comoving proton
density $n_{\rm com}$. However, since $\Delta_{\rm com} \simeq R/\Gamma$ and 
$n_{\rm com} \simeq n_{\rm o} \Gamma$, $\Delta_{\rm com} n_{\rm com} \simeq R n_{\rm o}$; 
hence we can express the {\sl dynamic} threshold in terms of the 
shock radius $R$ and ambiend density $n_{\rm o}$ as 

\begin{equation}
n_{\rm o} \,  \sigma_{_{\rm p \gamma}} R \gsim b \gamma ~~~{\rm or}
~~~ n_{\rm o}  \sigma_{_{\rm p \gamma}} R  \, \Gamma^2 \gsim 2 .
\label{dynamic}
\end{equation}
\

with the kinetic threshold used in the last step above.

It was noticed in \citep{km99} that the constraints on the thresholds given 
above can be alleviated if a fraction of the synchrotron photons (which are
only at a distance $R/\Gamma^2$ ahead of the RBW), can scatter in a ``mirror".
Scattering allows the RBW to catch-up with these photons, whose energy upon
reinterception by the RBW is now increased by a factor $4 \Gamma^2$, so that 
now the synchrotron photons have an energy $\epsilon^{\prime} = 4\, \Gamma^2 
\, b \, \gamma_e^2$. The kinematic threshold then becomes $\g_{_{\rm c}} \, 
\epsilon_s^{\prime} \gsim 2$ or 

\begin{equation}
b \, \Gamma^2 \, \gamma_{_{\rm c}}^3 \gsim \frac{1}{2} ~~ {\rm or} ~~ b \, 
\Gamma^5 \simeq \frac{1}{2} 
~~~~~~ \Gamma \gsim \left( \frac{1}{2b} \right)^{1/5} 
\label{kinematic}
\end{equation}
\

with the last steps assuming that the protons and electrons of the relativistic 
post-shock plasma have Lorentz factors $\gamma_{_{\rm c}} \simeq
\gamma_e \simeq \gamma_p \simeq \Gamma$.

The dynamic threshold condition remains the same, however when simplified using
Eq. (\ref{kinematic}) leads to 

\begin{equation}
n_{\rm o} \,  \sigma_{_{\rm p \gamma}} R \gsim b \gamma ~~~{\rm or}
~~~ n_{\rm o}  \sigma_{_{\rm p \gamma}} R  \, \Gamma^4 \gsim 2 .
\label{dynamic2}
\end{equation}
\

a condition a lot easier to fulfill than that of Eq. (\ref{dynamic}). 

\subsection{The Spectra: Scaling Arguments}

It is of interest to note that the fundamental radiative process of this model,
namely the $p\, \gamma \rightarrow e^+e^- -$ reaction, involves a threshold electron
energy [Eqs. (\ref{kinematic0}) or (\ref{kinematic})]. This is an important
fact because it implies the absence of electron injection at energies lower than 
$\g_{_{\rm c}}$ and leads naturally to a peak in the $\nu F_{\nu}$ GRB spectra at an 
(unspecified as yet) energy $E_{\rm p}$.  The presence of this peak energy
has been one of the landmark features of GRB spectra \citep{mallozzi95}, 
\citep{Preece00} and the issue of why $E_{\rm p} \sim m_ec^2$ has vexed scientists 
over the years. 

The more recent discovery of transients with 
values of $E_{\rm p}$ smaller than those found by {\sl BATSE} suggested that 
this may in fact be a selection effect with (perhaps) little importance.
However, the relation between $E_{\rm p}$ and the isotropic emitted energy
$E_{\rm iso}$ discovered by Amati et al. \citep{amati02} suggests that 
the value of $E_{\rm p}$ is not  random but it is tied to the physics of 
the burst and (as we content) to the conversion of its kinetic energy 
to radiation. 

The specificity of the above model allows the qualitative computation of 
the resulting specta in a rather straightforward fashion: The basic process for 
photon production is synchrotron radiation with corresponding energy at
$\epsilon_s \simeq b \, \Gamma^2$. The photons that are scattered on the 
``mirror", they have, upon their re-interception by the RBW, energy (on the 
RBW frame) $E \simeq b \, \Gamma^4$. These photons will then scatter: (a) elastically by 
``cold" electrons ($\gamma \sim 1$) of the RBW to preserve their energy at $\epsilon_1 
\simeq b \, \Gamma^4$, (b) inelastically by ``hot" ($\gamma \simeq \Gamma$)
electrons to boost their energy to $\epsilon_2 \simeq b \, \Gamma^6$ (or
rather to min$(b \, \Gamma^6, \Gamma)$ as the last scattering likely takes 
place in the Klein--Nishina regime). 
These energies will appear in the lab frame blue-shifted by a factor 
$\simeq \Gamma$ to occur correspondingly at energies $E_S \simeq b\,\Gamma^3,~
E_{BC} \simeq b \, \Gamma^5,~E_{IC}\simeq {\rm min}(b \, \Gamma^7, \Gamma^2)$.  
If the process operates 
close to its kimematic threshold, then, by virtue of Eq. (\ref{kinematic}), 
the energy of the second (bulk Comptonized) component will be at $E_{\rm p} 
\sim m_ec^2$, thereby 
resolving the issue raised by the {\sl BATSE} systematics \citep{mallozzi95}. In addition
to this aspect, this model also predicts the presence of additional peaks in 
the GRB spectra at energies $b \Gamma^3$ and ${\rm min}(\Gamma^2, b \Gamma^7)$
(all energies in units of $m_ec^2$).
For typical values of $\Gamma$ ($\sim 300$), these are respectively $E_S \simeq 
10$ eV and $E_{IC}\simeq 10-100$ GeV. 
The relative normalization of these three distinct components depends on the 
details of the processes involved \citep{mk06} (also \citep{kgm}) and 
imply the presence of prompt GRB emission in the optical and the {\sl GLAST}
energy bands.

\subsection{ Spectra and Variability: Detailed Calculations}

The properties of the model outlined above have been explored 
in detail through numerical simulations. To this end the 
distribution functions of the protons, electrons and photons have
been followed both in energy and time (or correspondingly in space), 
assuming a spherical shape for the emitting plasma. This approximation
is justified by considering that the radial width of the RBW only $\Delta \simeq 
R/\Gamma^2$ in the lab frame, has a comoving frame size $\Delta_{\rm com}\simeq R/
\Gamma$; since an external observer ``sees" only a fraction of the angular 
extent of the RBW of transverse spacial extent $R/\Gamma$ (of the same size 
as the photon horizon), the radiating plasma on the comoving frame, can be 
considered as spherical of radius $r \sim R/\Gamma \simeq \Delta_{\rm com}$. 
The calculations are thus preformed in the comoving frame and the radiation 
intensity is transformed to the lab frame at the very end.

The treatment (given in detail  in \citep{mk06}) follows closely that
of B\"ottcher \& Dermer \citep{BotDer98} AGN variability study. One important feature 
of this treatment is the fact that the radiation ``reflected" in the 
``mirror" contributes to the proton and electron losses only after 
the RBW photons have scattered on the ``mirror" and re-intercepted by the 
RBW. Therefore there is a causality constraint associated with the
corresponding terms in these equations. Assuming the ``mirror" to 
be extended, covering the range between $R_{\rm in}$ and $R_{\rm out}$, 
these terms become effective only after time $t_{\rm c} = (R_{\rm in}/c)\, 
2 \beta_{\Gamma}/(1+\beta_{\Gamma})$ where $\beta_{\Gamma}$ is the 
velocity of the RBW (normalized to $c$). 

\begin{figure}
\centering
\includegraphics[width=0.9\linewidth]{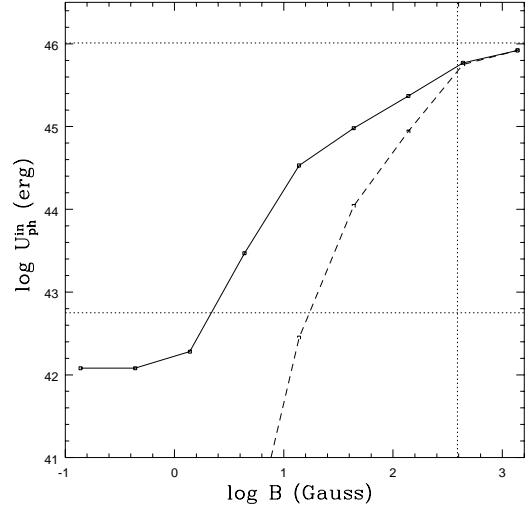}
\caption{The total energy radiated by the protons in a case with 
$n_p = n_e = 10^4 {\rm cm}^{-3}$ as a function of the magnetic field strength $B$ 
(both quantities measured in the comoving frame). The Lorentz factor of the RBW is 
$\Gamma = 
400$. Solid and dashed lines represent the cases with and without relativistic 
electrons
respectively. The lower horizontal dotted line is the total energy stored
in electrons while the upper one is the energy stored in protons. The 
vertical dotted line is the equipartition magnetic field.\label{UB}}
\end{figure}

The calculations at this first stage were performed assuming a constant 
value for the RBW Lorentz factor $\Gamma$. These were performed 
under two different assumptions concerning the presence or not of relativistic
electrons in the RBW, a feature crucial for the time development of 
the system: (a) No relativistic electrons initially present within the 
RBW. All necessary photons are produced (initially) from infinitesimal
(numerical) fluctuations in the photon field. However, these suffice to 
produce pairs which in turn produce the required photons as discussed 
above (see fig. 5 of \citep{mk06}). The increase in the ambient photons 
is exponential with characteristic time scale of the order of the light
crossing time of the shock; as shown in the said figure, the characteristic
time scale gets shorter the farther the parameters of system beyond their 
threshold values (i.e. the larger the value of the magnetic field $B$),
with the process eventually saturating when enough pairs have accumulated
to shield the protons from photon scattering. 
(b) Relativistic electrons are present in the RBW same in number and LF
as the protons. These cool very 
quickly to produce the photons necessary for the production of more 
electrons etc., as discussed above. The evolution in this case is much
faster and the proton energy can be radiated away far more efficiently than
in case (a) and for parameter values below those of the threshold.

These arguments are exemplified in Figure \ref{UB} where we plot the total energy 
radiated by the protons during the plasma traversal of a ``mirror" of Thompson 
depth $\tau_{\rm mir} =1$, $R_{\rm in} = 3 \, 10^{16}$ cm and $R_{\rm out} 
= 10^{17}$ cm. The presence of relativistic electrons facilitates the 
transfer of energy from the protons to the photons for a given value of 
$B$. For sufficiently large values of $B$, however, all the proton energy 
is eventually radiated away irrespective of the presence of the original 
relativistic electrons because the secondary pairs from the photon-proton 
collisions dominate the effects of the primary electrons. 

At the same time, because of the presence of the additional photons due 
to the electron cooling, the maximum available luminosity can be achieved
with parameter values (say that of the magnetic field $B$, below that 
necessitated by the threshold conditions. 

\begin{figure}
\centering
\includegraphics[width=0.9\linewidth]{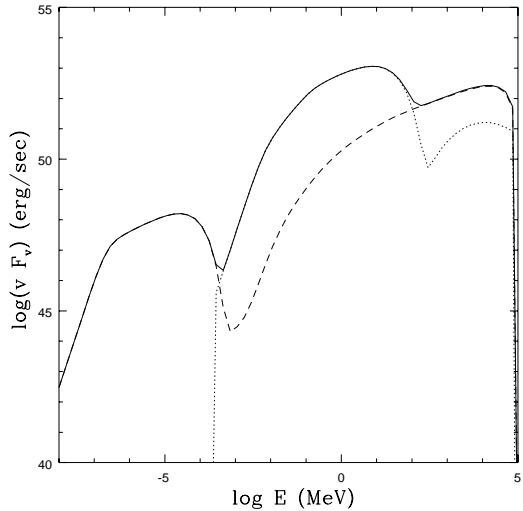}
\caption{The GRB spectrum for a case of $n_p = n_e = 10^5$ cm$^{-3}$ and $\Gamma = 
400$ and $B= 25$ G at maximum flux. The three components discussed in the text are
apparent with the middle one  having a peak near 1 MeV.\label{spectrum}}
\end{figure}

In Figure \ref{spectrum} we also plot the resulting spectrum at flux maximum for 
a RBW with $\Gamma = 400$, $B=25$ G and $n_p = n_e = 10^5$ cm$^{-3}$ (the values of 
magnetic field and the density are measured on the RBW frame). 
The spectrum exhibits clearly the three components 
discussed in \S 2.2, with the middle one at an energy $E_{\rm p}$ of a few MeV. A decrease 
in the value of the magnetic field will result in a lower value of $E_{\rm p}$, while
a decrease in the proton density $n_p$ (more correctly in the product $n_p \, 
\tau_{\rm mir}$) will lead to a decrease in the lumisosity of the bulk Comptonized
component relative to those of synchrotron and Inverse Compton ones. 

\section{The Afterglows}

\subsection{A Definition for the Afgerglow?}

Following their discovery \citep{Costa97}, afterglows have become an integral
part of the the GRB study and, in fact, they presently consume the largest fraction 
of effort to understand the GRB phenomenon. 

Based on the sparsely sampled early afterglow light curves and the ensuing 
theoretical models of their spectro-temporal behavior \citep{spn98, sph99},
their light curves were expected to be smooth with rather well specified breaks 
in their time profiles and spectra. 
The launch of {\sl Swift} and the possibility of continuously following a burst 
from the prompt to its afterglow stages, presented a number 
of surprises in  variance with the earlier models; a partial list includes 
(a) the steep decline in 
X-ray flux shortly after (or perhaps in the transition to the afterglow from) 
the prompt GRB phase (b) an unexpectedly slow decline following this steep 
early phase and (c) the delayed large flares in the X-ray band \citep{obrien06} 
thousands of seconds since the beginning of the burst. 

It appears, however, that in the excitement of the {\sl Swift} results the issue 
of precise definitions for the prompt and afterglow GRB stages and the transition 
from one to the other has been neglected. In this respect, the ``Supercritical Pile" 
model, precisely because it is founded on  well defined and specific physical processes, 
does offer an answer to the previous question. The remaining issue, then,  is to 
check whether this answer is in fact consistent with existing (or future) 
observations. 

The kinematic threshold associated with the model (Eq. \ref{kinematic}), in 
conjunction with the slowing down of the RBW as it accumulates mass from the surrounding 
medium, imply that the injection of electrons (and hence the GRB) should terminate 
as soon as $\Gamma$ drops below the value that satisfies the kinematic threshold 
condition. However, in the presence of an accelerated population of relativistic 
protons that extends to energies much higher than $\Gamma \, m_pc^2$, i.e. the 
energy associated with the (quasi-)thermal postshock population, the injection 
of relativistic electrons will continue even for $\Gamma \ll (2 / b)^{1/5}$;
this is because the the non-thermal ``tail" in the proton distribution guarantees 
the presence of protons with sufficiently high energies $\gamma_{_{\rm c}} \gg
 \Gamma$ which can fulfill the more general threshold condition $b \,
\gamma_{_{\rm c}}^3 \, \Gamma^2 \gsim 2$. 

Under these conditions the model predicts that electron injection will continue 
at a well defined energy but given in this case by the more general 
kinematic threshold of Eq. (\ref{kinematic0}) rather than at $\gamma_e = 
\Gamma$. Herein we put forward the {\sl (detector--sensitivity independent)} 
proposal that {\sl the transition from the prompt 
to the afterglow stage takes place when the RBW Lorentz factor drops below that of 
Eq. (\ref{kinematic}) or equivalently when $\Gamma <\gamma_{_{\rm c}}$.}

Using the arguments presented in \S 2.1, one can now estimate the 
energies at which the spectral peaks discussed there will occur 
under these new circumstances. Because most of the GRB phenomenology
covers the peak associated with the middle, bulk-Comptonized component
we restrict ourselves to this component. Following the arguments 
presented earlier, this peak will now be (in the lab frame) at an energy
 
\begin{equation}
E_{\rm p} \simeq b \gamma_{_{\rm c}}^2 \Gamma(t)^3 
\simeq 2 \left[ \frac{\Gamma(t)}{\gamma_{_{\rm c}}}\right] < 1
\label{Epeak2}
\end{equation}
\\

where the threhold condition $b \,
\gamma_{_{\rm c}}^3 \, \Gamma^2 \simeq 2$ 
has been used in the last step.

According to Eq. (\ref{Epeak2}) above, the value of the peak energy of the 
$\nu F_{\nu}$ spectra is time dependent and presumably decreases with decreasing 
$\Gamma(t)$. The decrease in time can be determined once a relation between
$\gamma_{_{\rm c}}$ and $\Gamma$ is found. Such a relation can be obtained 
from the threshold condition (Eq. \ref{kinematic0}) $b \, \gamma_{_{\rm c}}^3 
\, \Gamma^2 \gsim 2$ and it is  $\gamma_{_{\rm c}} \simeq [2/b \, 
\Gamma(t)^2]^{1/3}$. For magnetic field in equipartition with the postshock 
plasma $b \simeq 10^{-14} n_{\rm o}^{1/2} \Gamma(t)$, yielding $\gamma_{_{\rm c}} 
\simeq 6 \, 10^4 /[\Gamma(t) \, n_{\rm o}^{1/6}]$, where $n_{\rm o}$ is the 
ambient (preshock) density in cm$^{-3}$. Finally, the relation between $E_{\rm p}$ 
and $\Gamma(t)$ reads

\begin{equation}
E_{\rm p} 
\simeq 4.2 \, 10^{-2} \, n_{\rm o}^{1/6} \, \left[\frac{\Gamma(t)}{50}
\right]^2 ~.
\label{Epeak3}
\end{equation}
\

Therefore, one of the consequences of the electron injection process and spectrum 
formation within the present model is that the value of the energy of peak emission 
$E_{\rm p}$ should decrease with time as the 
GRB passes from the prompt emission to its afterglow stage. Eq. (\ref{Epeak3})
provides a specific dependence of this energy on $\Gamma$ and therefore
on time. For the ``standard" time dependence of $\Gamma$ for RBW propagation
through a medium of constant density ($\Gamma(t) \propto t^{-3/8}$ ) and for
magnetic field in equipartition, $E_{\rm p}(t) \propto t^{-3/4}$.
Clearly the late time evolution favors observations by detectors which 
are more sensitive to lower energies, as indeed seems to be the case with 
the {\sl Swift} observations. 

\begin{figure}
\centering
\includegraphics[width=1.1\linewidth]{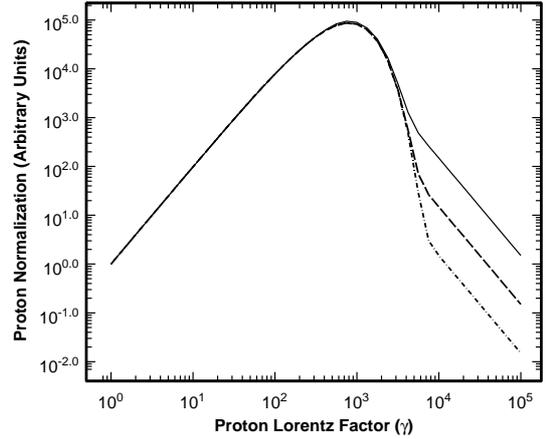}
\caption{The proton Distribution Function used in producing the afterglow
flux given in Figure \ref{flux}. Besides a relativistic Maxwellian of the form
$\gamma^2 \, e^{-\gamma/\Gamma}$ with $\Gamma = 400$, there is also a power law
`tail' to the spectrum with normalizations (relative to the peak of the Maxwellian)
0.1 (solid line), 0.01 (dashed line) and 0.001 (dot-dashed line). 
\label{ProtDF}}
\end{figure}

\subsection{Afterglow Light Curves}

One of the surprising  features of the GRB light curves obtained by {\sl Swift}
have been the steep decay of their X-ray flux, following the end of their 
prompt emission (or more specifically their becoming too faint for the BAT but 
not for the XRT). Generally, the models predicted flux decay $F_{\rm x} \propto 
t^{-1}$ \citep{spn98},\citep{sph99}, consistent with the observations of the earlier 
afterglows; steeper decrease in flux, attributed to emission from angles (between 
the fluid motion and the observer's line of sight) $\theta > 1/\Gamma$ 
\citep{kumpanait01} were also considered leading to X-ray flux profiles $F_{\rm x} 
\propto t^{-\alpha}, ~\alpha \gsim 2$. However, the new observations exhibited decay 
rates as steep as $t^{-6}$ \citep{obrien06}, in clear disagreement with the theory. 

\begin{figure}
\centering
\includegraphics[width=1.1\linewidth]{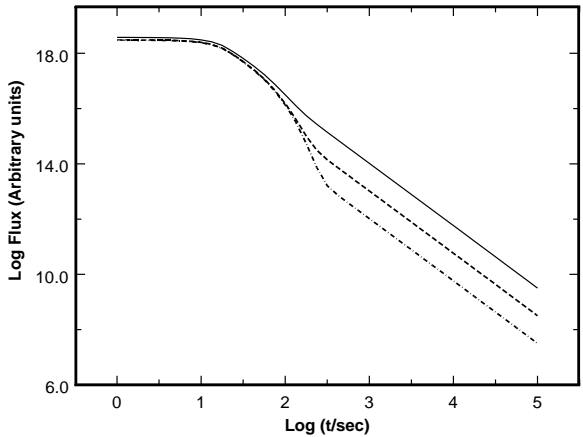}
\caption{ The synchrotron flux at energy $\epsilon = 10^{-6}$ as a function of 
time assuming $\Gamma(t) =$ constant for $t < 20$ seconds and $\Gamma(t) = (t/20)^{-3/8}$ 
for $t > 20$ seconds. The different curves correspond to the proton distributions with the 
power law normalizations given in Fig. \ref{ProtDF}. 
There is an apparent very steep drop of emission with time beginning at $t=20$ 
sec. \label{flux}}
\end{figure}

The model we have presented here incorporates the possibility of producing such 
steeply falling light curves. The essence behind such behavior within the present 
model lies in the form
of the proton distribution function behind the shock. Crucial for the extension of 
the model into the afterglow is the presence of a power-law ``tail" in addition
to the quasi-Maxwellian distribution of the bulk of the protons. Figure \ref{ProtDF}
shows the type of distribution we have been advocating in this section. We also 
assume that as the shock slows down the shape of the distribution function 
(i.e. the relative normalization of the power law and the Maxwellian) remains 
invariant and only shifts to lower energies. As the RBW slows down, the critical value 
of the proton energy necessary to fulfill the pair production threshold increases and 
as it sweeps past the maximum of the Maxwellian and into the power law section, the 
sharp drop in the proton density manifests as a sharp drop in the resulting photon flux. 
The corresponding change in the flux due to this effect is given in Figure \ref{flux},
with the different curves representing the flux produced by the corresponding proton
distribution of Figure \ref{ProtDF}. It is apparent that the model has sufficient freedom
to account for this sort of observation.

\section{GRB, XRR, XRF Unification?}

The discovery of transients with $E_{\rm p}$ at energies smaller than those of 
the classic GRB \citep{amati02} put in question the ``characteristic" value of 
$E_{\rm p}$ indicated by \citep{mallozzi95} raised the issue of their nature and 
their relation to the ``classic" GRB. This discovery and the restriction of the 
value of $E_{\rm p}$ within the ``Supercritical Pile" model to range near the 
electron rest mass seemed to effectively rule out the specific model (and for 
that matter to any model that would produce a limited range in the value of $E_{
\rm p}$).  

A resolution to this issue was suggested in \citep{ioka01}, who proposed
that the reduced value of $E_{\rm p}$ and the reduced peak flux of XRFs is related
to the angle of the observer relative to that of the velocity of the RBW, with the 
XRFs being observed at large such angles. This (and any similar) interpretation, 
however, tacitly assumes the presence of a ``unique" value for $E_{\rm p}\simeq 1$ 
for a RBW moving along the observer's line of sight, a feature that is not accounted
for by the typical GRB model. 

However, this particular account of the XRF - GRB unification has been challenged 
by the data of XRF 050416A \citep{mangano06}. The authors of this work pointed 
out that an orientation unification between 
the classic GRBs and XRFs implies that the source flux should increase as the 
RBW slows down since a larger fraction of the emitted luminosity would now
be emitted into the observer's line of sight. However, this particular event 
showed no such tendency, even though it was observed sufficiently early on so that 
this feature would not have been missed. Furthermore, it was shown that with an 
$E_{\rm p} \simeq 15$ keV, it nicely filled the gap between GRB and XRF in the 
Amati relation \citep{taka06}. 

The extension of the ``Supercritical Pile" model with the additional consideration of 
the high energy proton ``tail" suggested in the previous section can satisfy
both these above constraints: For an initial RBW Lorentz factor smaller than that implied
by the threshold condition of Eq. (\ref{kinematic}), the production of pairs 
will take place by the interaction of the photons not with the protons of the Maxwellian
part of the distribution but with the relativistic protons of its power law section. 
As indicated above, in this case, the value
of $E_{\rm p}$ will be smaller than the electron rest mass the lower $\Gamma(t)$ lies
below the threshold value necessary to pair produce with the protons of Lorentz
factor $\Gamma$ (i.e. those at the peak of the Maxwellian distribution). In fact, for field in equipartition with the 
protons, this value should be $E_{\rm p} \simeq 1 \, n_{\rm o}^{1/6} (\Gamma/250)^2 $, 
so that for $\Gamma \lsim 50$ this value will be in the range 20-40 keV, even for 
a RBW moving along the observer's line of sight. Also, because of the smaller
value of the RBW Lorentz factor $\Gamma$ the total luminosity and radiated energy
will be generally smaller, in agreement with the XRF observations. 

\section{The ``Mirror"}

One of the main aspects of the model described herein is that of the 
``mirror" whose purpose is to scatter the synchrotron photons so that 
they can be re-intercepted (blue-shifted now by $\sim 4\Gamma^2$) by the RBW. 
To this point it has been assumed that the albedo of this mirror is very
close to 1, this being the reason that it does not figure in the expressions 
of the thresholds (Eqs. \ref{kinematic},\ref{dynamic}). It is easy to see that
a smaller value for the mirror's albedo will not change the kinematic threshold
which is only concerned with the photon energies following the scattering 
by the mirror and the RBW. However, the dynamic threshold cares about the 
number of photons available to be scattered. Assuming that the fraction of 
the photons that scatter is proportional to the value of the ``mirror's" 
albedo $\tau_{\rm mir}$, then the corresponding expression for the dynamic 
threshold becomes

\begin{equation}
\tau_{\rm mir} n_{\rm o}  \sigma_{_{\rm p \gamma}} R  \, \Gamma^4 \gsim 1/2 
\label{dynamic3}
\end{equation}
\

The issue of the nature of the ``mirror" and the effect of its kinematic 
state on the thresholds has been discussed in detail in \citep{mk06}. 
One can easily see that an outlfowing ``mirror" will lead to an increase in the value 
of the LF of the RBW $\Gamma$ necessary to fulfill the kinematic threshold; 
however, the energy of the bulk Comptonized component of the resulting spectra 
will still appear, in the lab frame, at $E_{BC} \simeq 2 m_ec^2$ (assuming 
that the process operates near threshold), thus preserving this feature of the 
model. 

The simplest assumption concerning the ``mirror" is that it is simply due to Thompson 
scattering on the electrons of the circumburst matter (it appears unlikely that 
any matter would not be fully ionized in the intense GRB photon field). 
In this case, $\tau_{\rm mir} \simeq n_{\rm o}  \sigma_{_{\rm T}} R$, modifying the 
dynamic threshold to $\tau^2_{_{\rm T}} \Gamma^4 \gsim (\sigma_{_{\rm T}}/2
\sigma_{_{\rm p \gamma}})$, 
which for $n_{\rm o} \simeq 10^3$ cm$^{-3}$ and $R \simeq 3 10^{16}$ cm yields
$\Gamma \gsim 430$. These values of the density and radius are consistent with 
those encountered in a spherical wind of mass loss $\dot M \simeq 3 \, 10^{-6}$ 
M$_{\odot}$ yr$^{-1}$ and velocity $v \simeq 10^8$ cm/s, parameters consistent 
with those associated with the Wolf-Rayet stars whose collapse presumably leads to a GRB.

The process responsible for producing the radiation observed in the {\sl BATSE} 
and {\sl Swift} energy bands within this model, i.e. the bulk Comptonization of 
the synchrotron radiation reflected in the ``mirror", provides the possibility
of narrow `spikes' in the GRB light curves, the presence of lags between hard 
and soft photons in these spikes and allows for rough estimates of 
these timing features: The 
synchrotron photons emitted by the RBW are at a distance $\Delta \sim R/\Gamma^2$ 
ahead of it and their scattering in the ``mirror" produces (in the infinitely thin
mirror approximation) a photon shell of width $\Delta$; the bulk Comptonized 
component is produced as these photons are swept by the RBW; this ``sweeping" 
will take place on a time scale of order $\Delta t \sim \Delta/c \Gamma^2 \sim 
R/c \Gamma^4 \sim 10^{-4}$ sec (and proportionally longer by $\delta R/\Delta$
if the ``mirror" thickness $\delta R$ is larger than $\Delta$).
Therefore, the model can produce fine structure in the 
GRB light curves by a process which could be considered as a variant of that
of `internal shocks', while in reality involving only external ones. In 
addition, photons arriving to the observer at slightly different paths after
these repeated scatterings will have slightly different energies resulting 
to lags of the same order of magnitude, in general agreement with observations 
\citep{norris00}.

\section{Conclusions}

We have presented above the outline (and some details) of a model that
provides a well defined procedure for converting the kinetic energy of the 
RBW of GRB into relativistic electrons and then into photons. To the best
of our knowledge, this constitutes the first attempt to provide a detailed
model of the  dissipation (i.e. the conversion to radiation) of the kinetic 
energy of the GRB blast waves. In this respect the highly relativistic 
state of 
these flows is instrumental in effecting the dissipation of proton energy. 
Furthermore, the kinematic threshold of the reactions involved translates to 
a minimum energy in the injected electron distribution, which in turn leads
naturally to a peak in the $\nu F_{\nu}$ GRB spectra that lies (because of
the details of the same kinematics) in the vicinity of the electron rest mass,
in agreement with observation. 

The model provides also for a well defined demarkation between the prompt and 
the afterglow emissions based on the section of the proton distribution function
which contributes to the electron injection: The prompt emission terminates 
when the injection shifts (because of the decrease in the RGW Lorentz factor) 
from the (relativistic) Maxwellian to the power law section of that 
distribution. The kinematics of the pair production and spectral formation 
then imply that the energy of peak emission $E_{\rm p}$ shifts to lower
energies, also in general agreement with observation.

While our calculations todate have been limited to the $p \, \gamma
\rightarrow p \, e^-e^-$ reaction, the same photons could also produce
pions, whose radiation can be added to the calculations, which become 
somewhat more complicated but without fundamental changes of the main
point of the model. However, the possibility of photo-pion production
can lead to a qualitatively new process, namely the production of neutrinos 
of energies that can be easily calculated:
 A rough estimate of the neutrino energy is $\sim 5\%$ the 
enery of the proton; given that the protons of the RBW frame have energies
$E_{\rm p} \sim \Gamma \, m_pc^2$ and in the lab frame $E_{\rm p, lab} \sim
\Gamma^2 \, m_pc^2$, the lab neutrino energy will be $E_{\nu} \sim 8$
TeV $(\Gamma/400)^2$, a value of interest for the present and upcoming
neutrino telescopes. One should note that this estimate does not consider
the possibility of the power-law tail in the spectrum considered above. 
Should that be present, then correspondingly higher energies are possible.

Finally, this model can produce light curves with peaks of very narrow 
duration, despite the large size of the emission region, provided that 
the scattering medium has inhomogeneities of sufficiently short longitudinal
dimensions. Robust as it is, this model cannot (as yet at least) address
all questions pertaining to the physics of GRB, however, the GRB phenomenology
is sufficiently diverse that addressing a fraction only of the open issues
constitutes concrete progress.

\section*{Acknowledgments}

This work has been supported by the program Pythagoras
at the Univsrsity of Athens and an INTEGRAL GO grant
at GSFC and UMBC.


\end{document}